\RequirePackage{fix-cm}
\documentclass{svjour3}                     % onecolumn (standard format)
\smartqed  % flush right qed marks, e.g. at end of proof
\usepackage{graphicx}
\usepackage{amsmath}
\begin{document}

\title{Controlled remote state preparation via partially entangled quantum channel%\thanks{Grants or other notes
}


\author{ Chun Wang          \and
        Zhi Zeng  \and Xi-Han Li* %etc.
}

\authorrunning{C. Wang, Z. Zeng, X. H. Li} % if too long for running head

\institute{C. Wang, Z. Zeng, X. H. Li*\\
              Department of Physics, Chongqing University, Chongqing, China \\
              \email{xihanlicqu@gmail.com}            \\ \\
              X. H. Li\\Department of Physics and Computer Science, Wilfrid Laurier University, Waterloo, Canada
}

\date{Received: date / Accepted: date}

\maketitle

\begin{abstract}
We propose two controlled remote state preparation protocols via partially entangled channels. One prepares a single-qubit state and the other prepares a two-qubit state. Different from other controlled remote state preparation schemes which also utilize partially entangled channels, neither auxiliary qubits nor two-qubit unitary transformations are required in our schemes and the success probabilities are independent of the coefficients of the quantum channel. The success probabilities are 50\% and 25\% for arbitrary single-qubit states and two-qubit states, respectively. We also show that the success probabilities can reach 100\% for restricted classes of states.

% \PACS{03.67.Pp \and 03.67.Hk \and 03.65.Ud}
\end{abstract}

\section{Introduction}
Transmission of a quantum state that carries secret information is a crucial step in quantum communication. Other than physically sending the state through a noise channel,  there are two main communication methods each of which resorts to the pre-shared quantum entanglement and classical communication. One is the quantum teleportation \cite{tele} and the other is remote state preparation (RSP) \cite{RSP_lo}. In quantum teleportation, the sender Alice has the physical instance of the quantum state, which can be unknown to her. With the help of projective measurement with an entangled basis and classical communication, the unknown quantum state can be restored by a remote receiver Bob with some unitary operations. By contrast, in RSP protocols the sender only has the complete classical knowledge of the state, according to which she performs some positive operator valued measurement (POVM). Then the receiver can get the desired state with her measurement results. Although both quantum teleportation and RSP rely on quantum entanglement and classical communication, these two resources can be traded off against each other in RSP while in quantum teleportation they cannot \cite{RSP_Bennet}. We show that in the high-entanglement limit the asymptotic classical communication cost of remotely preparing a general qubit is one bit, which is half the corresponding cost in quantum teleportation \cite{RSP_Bennet}. In 2000, Pati demonstrated that a single-qubit state chosen from equatorial or polar great circles on a Bloch sphere can be remotely prepared with one classical bit from Alice to Bob if they share one Einstein-Podolsky-Rosen (EPR) pair in advance \cite{RSP_pati}. Subsequently, many novel RSP schemes were proposed \cite{low_entanglement_RSP,optimal_RSP,generalized_RSP,oblivious_RSP,faithful_RSP,cv_RSP} and experimental implementations were also reported \cite{RSPE1,RSPE2,RSPE3,RSPE4,RSPE5,RSPE6}.

In conventional RSP protocols, there is one sender who knows the information of the state to be prepared and one receiver who has no knowledge about the state. In 2007, Xia \emph{et al} proposed a novel multiparty remote state preparation scheme, in which two senders share the knowledge of the state. If and only if they agree to collaborate, the receiver can recover the quantum state \cite{JRSP_I_GC1}. Later, An \emph{et al} put forward a joint remote state preparation (JRSP) scheme for an arbitrary single-qubit state \cite{JRSP_I_GC2}. In this scheme, the knowledge of the state was divided into two parts, the amplitude information and the phase information, which are held by two separate senders. This division method was widely adopted by the subsequent JRSP schemes \cite{JRSP1,JRSP2,JRSP3}. The number of the senders was also extended to $M$ in Ref. \cite{AN1}. In 2010, An proposed new protocols for joint remote preparation via the W and W-type states \cite{AN2}. A protocol for joint remote state preparation of a W-type state was also presented in the same year \cite{AN3}.

 In addition to the JRSP, there is also another branch of RSP called controlled remote state preparation (CRSP), which transmits information in a controlled manner. In CRSP schemes, one or several controllers are introduced in addition to the sender and receiver. The difference between the controller in CRSP scheme and the added sender in JRSP protocol is that the controller has no information about the state to be prepared while the added sender has partial information about that.
 In 2007, Xiao \emph{et al} proposed a remote preparation protocol of a two-qubit entangled state via the W state channels, in which the original state can be prepared at either side of the two receivers who are distantly separated from each other \cite{CRSP_IIE_2WC3}. The implementation of this protocol was conditioned on the cooperation of both two receivers, so that we can view the two receivers as an actual receiver and a controller. In Ref. \cite{AN1}, a state preparation protocol which can only be realized with the permission of supervisors was also discussed.  In 2009, Wang \emph{et al} presented a scheme for remotely preparing a single-qubit state via the controls of many agents in a network, in which the name of controlled RSP was first brought up \cite{MCRSP_I_NE_EPR}. In their article, the agents' control was achieved by utilizing quantum key distribution and the scheme was also generalized to a class of $m$-qubit entangled state. In the same year, a scheme for multiparty-controlled remote preparation of a two-particle state by using two non-maximally GHZ states was proposed \cite{MCRSP_II_2GC3}, whose success probability achieved 50\% when using maximally entangled channel. Later, CRSP schemes of two-qubit states were proposed via different quantum channels, one single EPR pair and a GHZ state \cite{CRSP_II_E+G(C)}, a partially entangled tripartite GHZ state and a W-type state \cite{CRSP_II_GC+WC1,CRSP_II_GC+WC2} or a brown state \cite{CRSP_II_III_B}. In Ref. \cite{CRSP_II_III_B}, they also presented a CRSP scheme of a three-qubit state via the quantum channel composed of a Brown state and a Bell state.
Most recently, two controlled remote state preparation of an arbitrary single-qubit state schemes were presented by using the Affleck-Kennedy-Lieb-Tasaki (AKLT) state \cite{CRSP_I_A4}. In addition to the CRSP schemes which consider the qubit states, there were also CRSP protocols for qudit states \cite{CRSP_qudit1,CRSP_qudit2}. Moreover, remote state preparation schemes with several senders and controllers have been proposed, called controlled joint remote state preparation (CJRSP) schemes \cite{CJRSP_II_E,CJRSP_II_CC8,CJRSP_I}.

According to the degree of entanglement of the quantum channels, the CRSP schemes can be classed into two groups, one uses maximally entangled channels \cite{MCRSP_I_NE_EPR,CRSP_II_E+G(C),CRSP_II_III_B,CRSP_qudit2,CJRSP_II_E} and the other utilizes non-maximally entangled ones \cite{CRSP_IIE_2WC3,MCRSP_II_2GC3,CRSP_II_E+G(C),CRSP_II_GC+WC1,CRSP_II_GC+WC2,CRSP_qudit1,CJRSP_II_CC8,CJRSP_I}. The advantage of using maximally entangled channel is obvious. The optimal success probability and fidelity can be achieved with the help of the maximally entangled channel. However, it is not easy to generate and maintain the maximal entanglement in a practical environment. Although many methods have been proposed to distill a maximally entangled state from polluted states \cite{ep1,ep2,ep3,ep4,ep5,ep6,ep7,ep8,ep9,ep10}, many RSP schemes use the imperfect channels directly.
Inevitably, both the success probability and the fidelity of the state decrease with the degree of entanglement of the quantum channel. In most of those CRSP schemes that use non-maximally entangled channels, auxiliary qubits and two-qubit unitary transformations are required to prepare the target state. Then the CRSP schemes succeed with certain probabilities which depend on the degree of entanglement of the quantum channels.

In this letter, we propose two CRSP schemes via partially entangled channels, one for single-qubit states and the other for two-qubit states. We show that different from other CRSP schemes that also use partially entangled channels, our schemes do not require auxiliary particles and the success probabilities are independent of the quantum channels. The success probabilities are 50\% and 25\% for really arbitrary single-qubit and two-qubit states, respectively. And for some restricted classes of states the success probability of remote preparation reaches 100\% in principle.

\section{CRSP of a single-qubit state}
To remotely prepare a state in the receiver Bob's location controlled by Charlie, these three parties share the following partially entangled three-qubit state in advance.
\begin{eqnarray}
\vert \Psi\rangle_{ABC}=\frac{1}{\sqrt{2}}(\vert 000\rangle +a\vert 111\rangle+b\vert 110\rangle)_{ABC}.
\end{eqnarray}
Here $\vert a\vert ^2+\vert b \vert^2=1$. The subscript $A,B,C$ denote the particles belong to Alice, Bob and Charlie, respectively. This kind of states was discovered by Gao \emph{et al} in 2008 for perfect controlled teleportation \cite{gao}, which was also known as the maximal slice state in Ref. \cite{ms}.
Now we first illustrate the CRSP of arbitrary single-qubit state via the partially entangled channel and then we show the success probability can be 100\% for some specific states.

\subsection{CRSP of an arbitrary single-qubit state}

An arbitrary single-qubit state can be written as
\begin{eqnarray}
\vert \psi\rangle=\cos\theta \vert 0\rangle + e^{i\phi}\sin\theta\vert 1\rangle.
\end{eqnarray}
The sender Alice knows the complete information about this state, including the amplitude information $\theta$ and the phase information $\phi$. According to the information, Alice performs single-qubit measurement onto the orthogonal basis
\begin{eqnarray}
\vert \varphi_0\rangle&=&\cos\theta \vert 1\rangle - e^{i\phi}\sin\theta\vert 0\rangle,\nonumber\\
\vert \varphi_1\rangle&=&\cos\theta \vert 0\rangle +e^{-i\phi}\sin\theta\vert 1\rangle.
\end{eqnarray}
Then the state can be rewritten as
\begin{eqnarray}
\vert \Psi\rangle_{ABC}&=&\frac{1}{\sqrt{2}}\vert \varphi_0\rangle_A(a\cos\theta \vert 11\rangle+b\cos\theta \vert 10\rangle-e^{-i\phi}\sin\theta\vert 00\rangle)_{BC}\nonumber\\
&+&\frac{1}{\sqrt{2}}\vert \varphi_1\rangle_A(\cos\theta\vert 00\rangle+e^{i\phi}a\sin\theta \vert 11\rangle+e^{i\phi}b\sin\theta \vert 10\rangle)_{BC}. \label{1}
\end{eqnarray}
When Alice's measurement result is $\vert \varphi_0\rangle_A$, the CRSP fails since both the controller and the receiver have no information of $\phi$. They cannot restore the desired state without the phase information. If Alice's result is $\vert \varphi_1\rangle_A$, the state of Bob and Charlie collapses to corresponding state $\vert \Psi_1\rangle_{BC}=\frac{1}{\sqrt{2}}(\cos\theta\vert 00\rangle+e^{i\phi}a\sin\theta \vert 11\rangle+e^{i\phi}b\sin\theta \vert 10\rangle)_{BC}$. Then the controller Charlie measures his qubit  with the basis
\begin{eqnarray}
\vert \tau_+\rangle=\frac{1}{\sqrt{(1+b)^2+a^2}}[(1+b)\vert 0\rangle+a\vert 1\rangle],\nonumber\\
\vert \tau_-\rangle=\frac{1}{\sqrt{(1-b)^2+a^2}}[(1-b)\vert 0\rangle-a\vert 1\rangle].
\end{eqnarray}
The state can be written as
\begin{eqnarray}
\vert \Psi_1\rangle_{BC}= \frac{1+b}{\sqrt{(1+b)^2+a^2}}\vert \tau_+\rangle_C \otimes \vert \psi\rangle_B+ \frac{1-b}{\sqrt{(1-b)^2+a^2}}\vert \tau_-\rangle_C \otimes \sigma_z\vert \psi\rangle_B.
\end{eqnarray}
From the expression, we find Charlie's result and Bob's state have a corresponding relation. If Charlie allows the RSP procedure to take place, he sends his measurement results to Bob. Then Bob can get the desired state with unitary operation $I$ or $\sigma_z=\vert 0\rangle\langle0\vert -\vert 1\rangle\langle1\vert$ according to Charlie's information.

In conclusion, the total success probability is 50\% for the CRSP of an arbitrary single-qubit state via the partially entangled channel, which is the same as the CRSP scheme via maximally entangled channel \cite{MCRSP_I_NE_EPR}.

\subsection{CRSP of specific single-qubit states}
If we restrict the desired state to some subclasses, the success probability can be improved to 100\%. First of all, we suppose the state only contains amplitude information, i.e., a real qubit $\vert \psi^a\rangle=\cos\theta \vert 0\rangle + \sin\theta\vert 1\rangle$. We call it the amplitude state in the following text for simplicity. We set $\phi=0$ in the aforementioned procedure. In this case, Alice's measurement basis becomes
\begin{eqnarray}
\vert \varphi'_0\rangle&=&\cos\theta \vert 1\rangle - \sin\theta\vert 0\rangle,\nonumber\\
\vert \varphi'_1\rangle&=&\cos\theta \vert 0\rangle + \sin\theta\vert 1\rangle.
\end{eqnarray}
Therefore, Eq.(\ref{1}) can be rewritten as
\begin{eqnarray}
\vert \Psi\rangle_{ABC}&=&\frac{1}{\sqrt{2}}\vert \varphi'_0\rangle_A(a\cos\theta \vert 11\rangle+b\cos\theta \vert 10\rangle-\sin\theta\vert 00\rangle)_{BC}\nonumber\\
&+&\frac{1}{\sqrt{2}}\vert \varphi'_1\rangle_A(a\sin\theta \vert 11\rangle+b\sin\theta \vert 10\rangle+\cos\theta\vert 00\rangle)_{BC}
\end{eqnarray}
Then Charlie measures his qubit with $\vert \tau_\pm\rangle$ basis. If he agrees to cooperate and sends his results to Bob, Bob can get the desired state with a certain unitary operation according to Alice and Charlie's measurement results. The relation between Bob's recovery operation $U_B$, Alice's measurement results ($AM$) and Charlie's measurement outcomes ($CM$) is shown in Table \ref{tab1}. Here $\sigma_x=\vert 0\rangle\langle1\vert +\vert 1\rangle\langle0\vert$ and $\sigma_y=\vert 1\rangle\langle0\vert -\vert 0\rangle\langle1\vert$. It is obvious that the success probability is unit for the amplitude state.

\begin{table}[h]
\caption{The recovery operation $U_B$ conditioned on the measurement outcomes $AM$ of Alice and $CM$ of Charlie.}\label{tab1}
\begin{tabular}{cccl }
  \hline
   $AM$  & $CM$ & $U_B$  \\ \hline
  $\vert \varphi'_0\rangle_A$ & $\vert \tau_+\rangle_C$ &  $ \sigma_y$ \\
  $\vert \varphi'_0\rangle_A$ & $\vert \tau_-\rangle_C$ &  $ \sigma_x$ \\
  $\vert \varphi'_1\rangle_A$ & $\vert \tau_+\rangle_C$ &  $ I$ \\
  $\vert \varphi'_1\rangle_A$ & $\vert \tau_-\rangle_C$ &  $ \sigma_z$ \\
  \hline
\end{tabular}
\end{table}

Secondly, we come to the phase state which only contains the phase information as $\vert \psi^p\rangle=\frac{1}{\sqrt{2}}( \vert 0\rangle + e^{i\phi}\vert 1\rangle)$. In this case, Alice's measurement basis becomes
\begin{eqnarray}
\vert \varphi''_0\rangle&=&\vert 0\rangle + e^{-i\phi}\vert 1\rangle,\nonumber\\
\vert \varphi''_1\rangle&=& \vert 0\rangle -e^{-i\phi} \vert 1\rangle.
\end{eqnarray}
And Eq.(\ref{1}) can be rewritten as
\begin{eqnarray}
\vert \Psi\rangle_{ABC}&=&\frac{1}{\sqrt{2}}\vert \varphi''_0\rangle_A(\vert 00\rangle+e^{i\phi}a \vert 11\rangle+e^{i\phi}b\vert 10\rangle)_{BC}\nonumber\\
&&+\frac{1}{\sqrt{2}}\vert \varphi''_1\rangle_A(\vert 00\rangle-e^{i\phi}a \vert 11\rangle-e^{i\phi}b \vert 10\rangle)_{BC} \nonumber\\
&=&\frac{1}{\sqrt{2}}\frac{1+b}{\sqrt{(1+b)^2+a^2}}\vert \varphi''_0\rangle_A \vert \tau_+\rangle_C \otimes\vert \psi^p\rangle_B\nonumber\\&&+\frac{1}{\sqrt{2}} \frac{1-b}{\sqrt{(1-b)^2+a^2}}\vert \varphi''_0\rangle_A \vert \tau_-\rangle_C \otimes \sigma_z\vert \psi^p\rangle_B\nonumber\\&&+\frac{1}{\sqrt{2}}\frac{1+b}{\sqrt{(1+b)^2+a^2}}\vert \varphi''_1\rangle_A \vert \tau_+\rangle_C \otimes \sigma_z \vert \psi^p\rangle_B\nonumber\\&&+\frac{1}{\sqrt{2}} \frac{1-b}{\sqrt{(1-b)^2+a^2}}\vert \varphi''_1\rangle_A \vert \tau_-\rangle_C \otimes\vert \psi^p\rangle_B.
\end{eqnarray}

It is apparent that Bob can get the desired state with a $I$ or $\sigma_z$ operation according to Alice and Charlie's measurement results. The success probability is also 100\% in principle for the phase state.

\section{CRSP of a two-qubit state}
Now we discuss the remote preparation of two-qubit states. The three parties share two partially entangled tripartite states in advance.
\begin{eqnarray}
\vert \Psi\rangle_{A_1B_1C_1}=\frac{1}{\sqrt{2}}(\vert 000\rangle +a_1\vert 111\rangle+b_1\vert 110\rangle)_{A_1B_1C_1},\nonumber\\
\vert \Psi\rangle_{A_2B_2C_2}=\frac{1}{\sqrt{2}}(\vert 000\rangle +a_2\vert 111\rangle+b_2\vert 110\rangle)_{A_2B_2C_2}.
\end{eqnarray}
Here $\vert a_1\vert ^2+\vert b_1 \vert^2=1$ and $\vert a_2\vert ^2+\vert b_2 \vert^2=1$. The particles $A_1$,$A_2$ belong to the sender Alice, $B_1$,$B_2$ belong to the receiver Bob and $C_1$, $C_2$ are held by the controller Charlie. We demonstrate the CRSP of an arbitrary two-qubit state first and then discuss two specific cases.

\subsection{CRSP of an arbitrary two-qubit state}
The arbitrary two-qubit state can be written as
\begin{eqnarray}
\vert \psi\rangle=\alpha\vert 00\rangle +e^{i\phi_1}\beta\vert 01\rangle+ e^{i\phi_2}\delta\vert 10\rangle+e^{i\phi_3}\eta\vert 11 \rangle.
\end{eqnarray}
Here $\alpha, \beta,\delta$ and $\eta$ are all real coefficients and $\alpha^2+\beta^2+\delta^2+\eta^2=1$. The sender has the full information of the state to be remotely prepared. Firstly, Alice constructs a unitary matrix $U$ to establish a set of orthogonal measurement basis.
\begin{eqnarray}
V= \left(
    \begin{array}{cccc}
      \alpha & e^{-i\phi_1}\beta & e^{-i\phi_2}\delta & e^{-i\phi_3}\eta \\
      m\alpha & e^{-i\phi_1}m\beta & -e^{-i\phi_2}\delta/m & -e^{-i\phi_3}\eta/m \\
      e^{i\phi_1}\beta & -\alpha & e^{i\phi_3}\eta & -e^{i\phi_2}\delta \\
      e^{i\phi_1}m\beta & -m\alpha & -e^{i\phi_3}\eta/m & e^{i\phi_2}\delta/m \\
    \end{array}
  \right),
\end{eqnarray}
where $m=\sqrt{\delta^2+\eta^2}/\sqrt{\alpha^2+\beta^2}$. Alice's measurement basis is
\begin{eqnarray}
(\vert \varphi_{00}\rangle \ \ \vert \varphi_{01}\rangle \ \ \vert \varphi_{10}\rangle \ \ \vert \varphi_{11}\rangle)^T=V (\vert 00\rangle \ \ \vert 01\rangle \ \ \vert 10\rangle \ \ \vert 11\rangle)^T.
\end{eqnarray}
The superscript $T$ denotes the transpose of a vector or matrix. These vectors are mutually orthogonal to each other. Then the quantum channel can be rewritten in terms of Alice's measurement basis as
\begin{eqnarray}
&&\vert \Psi\rangle_{A_1B_1C_1} \otimes \vert \Psi\rangle_{A_2B_2C_2}\nonumber\\&=&\frac{1}{2}(\vert \varphi_{00}\rangle_{A_1A_2} \vert \Xi_{00}\rangle_{B_1C_1B_2C_2}+\vert \varphi_{01}\rangle_{A_1A_2} \vert \Xi_{01}\rangle_{B_1C_1B_2C_2}\nonumber\\&&+\vert \varphi_{10}\rangle_{A_1A_2} \vert \Xi_{10}\rangle_{B_1C_1B_2C_2}+\vert \varphi_{11}\rangle_{A_1A_2} \vert \Xi_{11}\rangle_{B_1C_1B_2C_2}).
\end{eqnarray}
Where
\begin{eqnarray}
&&\vert \Xi_{00}\rangle_{B_1C_1B_2C_2}\nonumber\\&=& [\alpha \vert 0000\rangle +e^{i\phi_1}\beta(a_2\vert 0011\rangle +b_2\vert 0010\rangle)+e^{i\phi_2}\delta(a_1\vert 1100\rangle +b_1\vert 1000\rangle)\nonumber\\&&+e^{i\phi_3}\eta(a_1a_2\vert 1111\rangle+a_1b_2\vert 1110\rangle+b_1a_2\vert 1011\rangle+b_1b_2\vert 1010\rangle)]_{B_1C_1B_2C_2},
\end{eqnarray}
\begin{eqnarray}
&&\vert \Xi_{01}\rangle_{B_1C_1B_2C_2}\nonumber\\&=& [m\alpha \vert 0000\rangle +e^{i\phi_1}m\beta(a_2\vert 0011\rangle +b_2\vert 0010\rangle)-e^{i\phi_2}\delta/m(a_1\vert 1100\rangle +b_1\vert 1000\rangle)\nonumber\\&&-e^{i\phi_3}\eta/m(a_1a_2\vert 1111\rangle+a_1b_2\vert 1110\rangle+b_1a_2\vert 1011\rangle+b_1b_2\vert 1010\rangle)]_{B_1C_1B_2C_2},
\end{eqnarray}
\begin{eqnarray}
&&\vert \Xi_{10}\rangle_{B_1C_1B_2C_2}\nonumber\\&=& [e^{-i\phi_1}\beta \vert 0000\rangle -\alpha(a_2\vert 0011\rangle +b_2\vert 0010\rangle)+e^{-i\phi_3}\eta(a_1\vert 1100\rangle +b_1\vert 1000\rangle)\nonumber\\&&-e^{-i\phi_2}\delta(a_1a_2\vert 1111\rangle+a_1b_2\vert 1110\rangle+b_1a_2\vert 1011\rangle+b_1b_2\vert 1010\rangle)]_{B_1C_1B_2C_2},
\end{eqnarray}
\begin{eqnarray}
&&\vert \Xi_{11}\rangle_{B_1C_1B_2C_2}\nonumber\\&=& [e^{-i\phi_1}m\beta \vert 0000\rangle -m\alpha(a_2\vert 0011\rangle +b_2\vert 0010\rangle)-e^{-i\phi_3}\eta/m(a_1\vert 1100\rangle +b_1\vert 1000\rangle)\nonumber\\&&+e^{-i\phi_2}\delta/m(a_1a_2\vert 1111\rangle+a_1b_2\vert 1110\rangle+b_1a_2\vert 1011\rangle+b_1b_2\vert 1010\rangle)]_{B_1C_1B_2C_2}.
\end{eqnarray}
Since Bob and Charlie have no information about the target state, they can not rebuild it from $\vert \Xi_{01}\rangle_{B_1C_1B_2C_2}$,$\vert \Xi_{10}\rangle_{B_1C_1B_2C_2}$ and $\vert \Xi_{11}\rangle_{B_1C_1B_2C_2}$. Only when Alice's measurement result is $\vert \Xi_{00}\rangle_{B_1C_1B_2C_2}$, Charlie performs two single-qubit measurement on $C_1$ and $C_2$ with basis $\vert \tau_{1\pm} \rangle$ and  $\vert \tau_{2\pm} \rangle$, respectively.
\begin{eqnarray}
\vert \tau_{j+}\rangle=\frac{1}{\sqrt{(1+b_j)^2+a_j^2}}[(1+b_j)\vert 0\rangle+a_j\vert 1\rangle],\nonumber\\
\vert \tau_{j-}\rangle=\frac{1}{\sqrt{(1-b_j)^2+a_j^2}}[(1-b_j)\vert 0\rangle-a_j\vert 1\rangle].
\end{eqnarray}
Here $j=1,2$. The relation between Charlie's measurement results ($CM$), Bob's collapsed state and the unitary operation ($ U_{B_1}\otimes U_{B_2}$) to get the desired state is shown in Table \ref{tab2}.
\begin{table}[h]
\caption{The relation between Charlie's measurement results, Bob's collapsed state and the unitary operations to recover the target state.}\label{tab2}
\begin{tabular}{cccl }
  \hline
   $CM$  & $State$ & $ U_{B_1}\otimes U_{B_2}$  \\ \hline
  $\vert \tau_+\rangle_{C_1}\vert \tau_+\rangle_{C_2}$ & $\alpha\vert 00\rangle +e^{i\phi_1}\beta\vert 01\rangle+ e^{i\phi_2}\delta\vert 10\rangle+e^{i\phi_3}\eta\vert 11 \rangle_{B_1B_2}$ &  $ I \otimes I$ \\
  $\vert \tau_+\rangle_{C_1}\vert \tau_-\rangle_{C_2}$ & $\alpha\vert 00\rangle -e^{i\phi_1}\beta\vert 01\rangle+ e^{i\phi_2}\delta\vert 10\rangle-e^{i\phi_3}\eta\vert 11 \rangle_{B_1B_2}$ &  $ I\otimes \sigma_z$ \\
  $\vert \tau_-\rangle_{C_1}\vert \tau_+\rangle_{C_2}$ & $\alpha\vert 00\rangle +e^{i\phi_1}\beta\vert 01\rangle- e^{i\phi_2}\delta\vert 10\rangle-e^{i\phi_3}\eta\vert 11 \rangle_{B_1B_2}$ &  $ \sigma_z \otimes I$ \\
  $\vert \tau_-\rangle_{C_1}\vert \tau_-\rangle_{C_2}$ & $\alpha\vert 00\rangle -e^{i\phi_1}\beta\vert 01\rangle- e^{i\phi_2}\delta\vert 10\rangle+e^{i\phi_3}\eta\vert 11 \rangle_{B_1B_2}$ &  $ \sigma_z \otimes \sigma_z$ \\
  \hline
\end{tabular}
\end{table}
With Charlie's collaboration, Bob can get the desired state with success probability 25\%. It is also the same as that of the CRSP scheme which uses maximally entangled channel \cite{CRSP_II_E+G(C)}.

\subsection{CRSP of restricted two-qubit states}
If we limit the state to some specific subclasses, the success probability can be improved to 100\%. The first kind of states only carriers the amplitude information $\vert \psi^a\rangle=\alpha\vert 00\rangle +\beta\vert 01\rangle+ \delta\vert 10\rangle+\eta\vert 11 \rangle.$ In this case, Alice's measurement basis can be written as
\begin{eqnarray}
 \left(
    \begin{array}{c}
       |\varphi'_{00}\rangle \\
       |\varphi'_{01}\rangle \\
       |\varphi'_{10}\rangle \\
       |\varphi'_{11}\rangle \\
    \end{array}
    \right)
    =\left(
       \begin{array}{cccc}
      \alpha & \beta & \delta & \eta \\
      \beta & -\alpha & \eta & -\delta \\
      \delta & -\eta & -\alpha & \beta \\
      -\eta & -\delta & \beta & \alpha \\
  \end{array}
  \right)
  \left(
         \begin{array}{c}
           |00\rangle \\
           |01\rangle \\
           |10\rangle \\
           |11\rangle \\
         \end{array}
       \right).
\end{eqnarray}
Charlie also measures his particles $C_1$ and $C_2$ with $\vert \tau_{1\pm}\rangle$ and $\vert \tau_{2\pm}\rangle$ bases, respectively. With Alice and Charlie's information, Bob can always get the desired state. The relation between Alice's measurement results (AM), Charlie's measurement result (CM), Bob's state and the corresponding unitary operations ($U_{B_1}\otimes U_{B_2}$) is shown in Table \ref{tab3}. We find that no matter what results Alice and Charlie get, Bob can always obtain the desired state. The success probability of the CRSP scheme of a real coefficients two-qubit state achieves 100\% in principle, higher than that of other CRSP schemes which also utilize partially entangled channels \cite{MCRSP_II_2GC3,CRSP_II_GC+WC1,CRSP_II_GC+WC2}.

\begin{table}[h]
\caption{The relation between Alice's measurement results ($AM$), Charlie's measurement results ($CM$), Bob's state and the unitary operations in the CRSP protocol of the two-qubit amplitude state.}\label{tab3}
\begin{tabular}{cccl }
  \hline
   $AM$  & $CM$ & $state$ & $ U_{B_1}\otimes U_{B_2}$  \\ \hline
   $\vert \varphi'_{00}\rangle_{A_1A_2}$  & $\vert \tau_{1+}\rangle_{C_1}\vert \tau_{2+}\rangle_{C_2}$
   &$(\alpha\vert 00\rangle +\beta\vert 01\rangle+ \delta\vert 10\rangle+\eta\vert 11 \rangle)_{B_1B_2}$& $ I \otimes I$  \\
   $\vert \varphi'_{00}\rangle_{A_1A_2}$  & $\vert \tau_{1+}\rangle_{C_1}\vert \tau_{2-}\rangle_{C_2}$
   &$(\alpha\vert 00\rangle -\beta\vert 01\rangle+ \delta\vert 10\rangle-\eta\vert 11 \rangle)_{B_1B_2}$& $ I \otimes \sigma_z$  \\
   $\vert \varphi'_{00}\rangle_{A_1A_2}$  & $\vert \tau_{1-}\rangle_{C_1}\vert \tau_{2+}\rangle_{C_2}$
   &$(\alpha\vert 00\rangle +\beta\vert 01\rangle- \delta\vert 10\rangle-\eta\vert 11 \rangle)_{B_1B_2}$& $ \sigma_z \otimes I$  \\
   $\vert \varphi'_{00}\rangle_{A_1A_2}$  & $\vert \tau_{1-}\rangle_{C_1}\vert \tau_{2-}\rangle_{C_2}$
   &$(\alpha\vert 00\rangle -\beta\vert 01\rangle- \delta\vert 10\rangle+\eta\vert 11 \rangle)_{B_1B_2}$& $ \sigma_z \otimes \sigma_z$  \\
   $\vert \varphi'_{01}\rangle_{A_1A_2}$  & $\vert \tau_{1+}\rangle_{C_1}\vert \tau_{2+}\rangle_{C_2}$
   &$(\beta\vert 00\rangle -\alpha\vert 01\rangle+ \eta\vert 10\rangle-\delta\vert 11 \rangle)_{B_1B_2}$& $ I \otimes \sigma_y$  \\
   $\vert \varphi'_{01}\rangle_{A_1A_2}$  & $\vert \tau_{1+}\rangle_{C_1}\vert \tau_{2-}\rangle_{C_2}$
   &$(\beta\vert 00\rangle +\alpha\vert 01\rangle+ \eta\vert 10\rangle+\delta\vert 11 \rangle)_{B_1B_2}$& $ I \otimes \sigma_x$  \\
   $\vert \varphi'_{01}\rangle_{A_1A_2}$  & $\vert \tau_{1-}\rangle_{C_1}\vert \tau_{2+}\rangle_{C_2}$
   &$(\beta\vert 00\rangle -\alpha\vert 01\rangle- \eta\vert 10\rangle+\delta\vert 11 \rangle)_{B_1B_2}$& $ \sigma_z \otimes \sigma_y$  \\
   $\vert \varphi'_{01}\rangle_{A_1A_2}$  & $\vert \tau_{1-}\rangle_{C_1}\vert \tau_{2-}\rangle_{C_2}$
   &$(\beta\vert 00\rangle +\alpha\vert 01\rangle- \eta\vert 10\rangle-\delta\vert 11 \rangle)_{B_1B_2}$& $ \sigma_z \otimes \sigma_x$  \\
   $\vert \varphi'_{10}\rangle_{A_1A_2}$  & $\vert \tau_{1+}\rangle_{C_1}\vert \tau_{2+}\rangle_{C_2}$
   &$(\delta\vert 00\rangle -\eta\vert 01\rangle- \alpha\vert 10\rangle+\beta\vert 11 \rangle)_{B_1B_2}$& $\sigma_y\otimes\sigma_z $  \\
   $\vert \varphi'_{10}\rangle_{A_1A_2}$  & $\vert \tau_{1+}\rangle_{C_1}\vert \tau_{2-}\rangle_{C_2}$
   &$(\delta\vert 00\rangle +\eta\vert 01\rangle- \alpha\vert 10\rangle-\beta\vert 11 \rangle)_{B_1B_2}$& $ \sigma_y\otimes I$  \\
   $\vert \varphi'_{10}\rangle_{A_1A_2}$  & $\vert \tau_{1-}\rangle_{C_1}\vert \tau_{2+}\rangle_{C_2}$
   &$(\delta\vert 00\rangle -\eta\vert 01\rangle+ \alpha\vert 10\rangle-\beta\vert 11 \rangle)_{B_1B_2}$& $\sigma_x\otimes \sigma_z$  \\
   $\vert \varphi'_{10}\rangle_{A_1A_2}$  & $\vert \tau_{1-}\rangle_{C_1}\vert \tau_{2-}\rangle_{C_2}$
   &$(\delta\vert 00\rangle +\eta\vert 01\rangle+ \alpha\vert 10\rangle+\beta\vert 11 \rangle)_{B_1B_2}$& $\sigma_x\otimes I $  \\
   $\vert \varphi'_{11}\rangle_{A_1A_2}$  & $\vert \tau_{1+}\rangle_{C_1}\vert \tau_{2+}\rangle_{C_2}$
   &$(\eta\vert 00\rangle +\delta\vert 01\rangle- \beta\vert 10\rangle-\alpha\vert 11 \rangle)_{B_1B_2}$& $\sigma_y\otimes \sigma_x $  \\
   $\vert \varphi'_{11}\rangle_{A_1A_2}$  & $\vert \tau_{1+}\rangle_{C_1}\vert \tau_{2-}\rangle_{C_2}$
   &$(\eta\vert 00\rangle -\delta\vert 01\rangle- \beta\vert 10\rangle+\alpha\vert 11 \rangle)_{B_1B_2}$& $ \sigma_y\otimes \sigma_y$  \\
   $\vert \varphi'_{11}\rangle_{A_1A_2}$  & $\vert \tau_{1-}\rangle_{C_1}\vert \tau_{2+}\rangle_{C_2}$
   &$(\eta\vert 00\rangle +\delta\vert 01\rangle+ \beta\vert 10\rangle+\alpha\vert 11 \rangle)_{B_1B_2}$& $ \sigma_x\otimes \sigma_x$  \\
   $\vert \varphi'_{11}\rangle_{A_1A_2}$  & $\vert \tau_{1-}\rangle_{C_1}\vert \tau_{2-}\rangle_{C_2}$
   &$(\eta\vert 00\rangle -\delta\vert 01\rangle+ \beta\vert 10\rangle-\alpha\vert 11 \rangle)_{B_1B_2}$& $\sigma_x\otimes\sigma_y $  \\
  \hline
\end{tabular}
\end{table}

The second kind of restricted states only carries the phase information, i.e., $\vert \psi^p\rangle=\vert 00\rangle +e^{i\phi_1}\vert 01\rangle+ e^{i\phi_2}\vert 10\rangle+e^{i\phi_3}\vert 11 \rangle.$ In this case, Alice's orthogonal measurement basis can be chosen as
\begin{eqnarray}
 \left(
    \begin{array}{c}
       |\varphi''_{00}\rangle \\
       |\varphi''_{01}\rangle \\
       |\varphi''_{10}\rangle \\
       |\varphi''_{11}\rangle \\
    \end{array}
    \right)
    =\left(
       \begin{array}{cccc}
      1 & e^{-i\phi_1} & e^{-i\phi_2} & e^{-i\phi_3} \\
      1 & e^{-i\phi_1} & -e^{-i\phi_2} & -e^{-i\phi_3} \\
      1 & -e^{-i\phi_1} & e^{-i\phi_2} & -e^{-i\phi_3} \\
      1 & -e^{-i\phi_1} & -e^{-i\phi_2} & e^{-i\phi_3} \\
  \end{array}
  \right)
  \left(
         \begin{array}{c}
           |00\rangle \\
           |01\rangle \\
           |10\rangle \\
           |11\rangle \\
         \end{array}
       \right).
\end{eqnarray}

The controller Charlie also does the same job. If Charlie agrees the procedure to continue, he tells Bob his measurement results. With Alice and Charlie's information, Bob can obtain the desired state by unitary operations. The details are shown in Table \ref{tab4}.  The success probability of this scheme is also 100\% in total.

\begin{table}[h]
\caption{The relation between Alice's measurement results ($AM$), Charlie's measurement results ($CM$), Bob's state and the unitary operations to get the target state in the CRSP protocol of the two-qubit phase state.}\label{tab4}
\begin{tabular}{cccl }
  \hline
   $AM$  & $CM$ & $state$ & $ U_{B_1}\otimes U_{B_2}$  \\ \hline
   $\vert \varphi''_{00}\rangle_{A_1A_2}$  & $\vert \tau_{1+}\rangle_{C_1}\vert \tau_{2+}\rangle_{C_2}$
   &$(\vert 00\rangle +e^{i\phi_1}\vert 01\rangle+ e^{i\phi_2}\vert 10\rangle+e^{i\phi_3}\vert 11 \rangle)_{B_1B_2}$& $ I \otimes I$  \\
   $\vert \varphi''_{00}\rangle_{A_1A_2}$  & $\vert \tau_{1+}\rangle_{C_1}\vert \tau_{2-}\rangle_{C_2}$
   &$(\vert 00\rangle -e^{i\phi_1}\vert 01\rangle+ e^{i\phi_2}\vert 10\rangle-e^{i\phi_3}\vert 11 \rangle)_{B_1B_2}$& $ I \otimes \sigma_z$  \\
   $\vert \varphi''_{00}\rangle_{A_1A_2}$  & $\vert \tau_{1-}\rangle_{C_1}\vert \tau_{2+}\rangle_{C_2}$
   &$(\vert 00\rangle +e^{i\phi_1}\vert 01\rangle- e^{i\phi_2}\vert 10\rangle-e^{i\phi_3}\vert 11 \rangle)_{B_1B_2}$& $ \sigma_z \otimes I$  \\
   $\vert \varphi''_{00}\rangle_{A_1A_2}$  & $\vert \tau_{1-}\rangle_{C_1}\vert \tau_{2-}\rangle_{C_2}$
   &$(\vert 00\rangle -e^{i\phi_1}\vert 01\rangle- e^{i\phi_2}\vert 10\rangle+e^{i\phi_3}\vert 11 \rangle)_{B_1B_2}$& $ \sigma_z \otimes \sigma_z$  \\
   $\vert \varphi''_{01}\rangle_{A_1A_2}$  & $\vert \tau_{1+}\rangle_{C_1}\vert \tau_{2+}\rangle_{C_2}$
   &$(\vert 00\rangle +e^{i\phi_1}\vert 01\rangle- e^{i\phi_2}\vert 10\rangle-e^{i\phi_3}\vert 11 \rangle)_{B_1B_2}$& $ \sigma_z \otimes I$  \\
   $\vert \varphi''_{01}\rangle_{A_1A_2}$  & $\vert \tau_{1+}\rangle_{C_1}\vert \tau_{2-}\rangle_{C_2}$
   &$(\vert 00\rangle -e^{i\phi_1}\vert 01\rangle- e^{i\phi_2}\vert 10\rangle+e^{i\phi_3}\vert 11 \rangle)_{B_1B_2}$& $ \sigma_z \otimes \sigma_z$  \\
   $\vert \varphi''_{01}\rangle_{A_1A_2}$  & $\vert \tau_{1-}\rangle_{C_1}\vert \tau_{2+}\rangle_{C_2}$
   &$(\vert 00\rangle +e^{i\phi_1}\vert 01\rangle+ e^{i\phi_2}\vert 10\rangle+e^{i\phi_3}\vert 11 \rangle)_{B_1B_2}$& $ I \otimes I$  \\
   $\vert \varphi''_{01}\rangle_{A_1A_2}$  & $\vert \tau_{1-}\rangle_{C_1}\vert \tau_{2-}\rangle_{C_2}$
   &$(\vert 00\rangle -e^{i\phi_1}\vert 01\rangle+ e^{i\phi_2}\vert 10\rangle-e^{i\phi_3}\vert 11 \rangle)_{B_1B_2}$& $I \otimes \sigma_z$  \\
   $\vert \varphi''_{10}\rangle_{A_1A_2}$  & $\vert \tau_{1+}\rangle_{C_1}\vert \tau_{2+}\rangle_{C_2}$
   &$(\vert 00\rangle -e^{i\phi_1}\vert 01\rangle+e^{i\phi_2}\vert 10\rangle-e^{i\phi_3}\vert 11 \rangle)_{B_1B_2}$& $I\otimes\sigma_z $  \\
   $\vert \varphi''_{10}\rangle_{A_1A_2}$  & $\vert \tau_{1+}\rangle_{C_1}\vert \tau_{2-}\rangle_{C_2}$
   &$(\vert 00\rangle +e^{i\phi_1}\vert 01\rangle+e^{i\phi_2}\vert 10\rangle+e^{i\phi_3}\vert 11 \rangle)_{B_1B_2}$& $ I\otimes I$  \\
   $\vert \varphi''_{10}\rangle_{A_1A_2}$  & $\vert \tau_{1-}\rangle_{C_1}\vert \tau_{2+}\rangle_{C_2}$
   &$(\vert 00\rangle -e^{i\phi_1}\vert 01\rangle-e^{i\phi_2}\vert 10\rangle+e^{i\phi_3}\vert 11 \rangle)_{B_1B_2}$& $\sigma_z\otimes \sigma_z$  \\
   $\vert \varphi''_{10}\rangle_{A_1A_2}$  & $\vert \tau_{1-}\rangle_{C_1}\vert \tau_{2-}\rangle_{C_2}$
   &$(\vert 00\rangle +e^{i\phi_1}\vert 01\rangle-e^{i\phi_2}\vert 10\rangle-e^{i\phi_3}\vert 11 \rangle)_{B_1B_2}$& $\sigma_z\otimes I $  \\
   $\vert \varphi''_{11}\rangle_{A_1A_2}$  & $\vert \tau_{1+}\rangle_{C_1}\vert \tau_{2+}\rangle_{C_2}$
   &$(\vert 00\rangle -e^{i\phi_1}\vert 01\rangle-e^{i\phi_2}\vert 10\rangle+e^{i\phi_3}\vert 11 \rangle)_{B_1B_2}$& $\sigma_z\otimes \sigma_z $  \\
   $\vert \varphi''_{11}\rangle_{A_1A_2}$  & $\vert \tau_{1+}\rangle_{C_1}\vert \tau_{2-}\rangle_{C_2}$
   &$(\vert 00\rangle +e^{i\phi_1}\vert 01\rangle-e^{i\phi_2}\vert 10\rangle-e^{i\phi_3}\vert 11 \rangle)_{B_1B_2}$& $ \sigma_z\otimes I$  \\
   $\vert \varphi''_{11}\rangle_{A_1A_2}$  & $\vert \tau_{1-}\rangle_{C_1}\vert \tau_{2+}\rangle_{C_2}$
   &$(\vert 00\rangle -e^{i\phi_1}\vert 01\rangle+e^{i\phi_2}\vert 10\rangle-e^{i\phi_3}\vert 11 \rangle)_{B_1B_2}$& $ I\otimes \sigma_z$  \\
   $\vert \varphi''_{11}\rangle_{A_1A_2}$  & $\vert \tau_{1-}\rangle_{C_1}\vert \tau_{2-}\rangle_{C_2}$
   &$(\vert 00\rangle +e^{i\phi_1}\vert 01\rangle+e^{i\phi_2}\vert 10\rangle+e^{i\phi_3}\vert 11 \rangle)_{B_1B_2}$& $I\otimes I $  \\
  \hline
\end{tabular}
\end{table}

\section{Discussion and summary}
We have demonstrated the CRSP schemes for a single-qubit state and a two-qubit state via the partially entangled channels. The success probability can be 100\% for states from restricted classes. And for arbitrary states, the success probability is 50\% and 25\% for single-qubit states and two-qubit states, respectively. In the single-qubit scheme, the sender and the controller each has to transmit 1 bit classical information to the receiver. From the viewpoint of the sender, this is half of classical information required in a controlled teleportation scheme of a single-qubit state \cite{ct}. In the two-qubit scheme, both of the sender and the controller have to send 2 bits classical information while 4 bits classical information are transmitted from the sender to the receiver in the quantum teleportation protocol of two-qubit states \cite{ctele}.  The sender's information can be reduced to 1 bit in the CRSP scheme of arbitrary two-qubit states. Since only one of her four measurement results is useful, Alice can only say "yes" or "no" to tell Bob whether he has the chance to recover the state.

Our schemes can be generalized to prepare $N$-qubit states via $N$ such partially entangled tripartite channels. Each of the three parties holds $N$ particles. Alice performs $N$-qubit measurement onto the basis established with the information of state. Charlie implements $N$ single-qubit measurements with $\vert \tau_\pm\rangle$ basis. The receiver can get the desired state with some unitary operations according to Alice and Charlie's measurement results. It is not difficult to conjecture that for an arbitrary $N$-qubit state, the success probability is $1/2^N$. With our partially entangled channel, the success probability can be can be 100\% for some restricted $N$-qubit states, i.e., the states only carry one kind of information, the amplitude or the phase information.
In a word, the success probabilities of our CRSP schemes utilizing partially entangled channel are  the same as that of schemes which use the maximally entangled channels.

Here we have to point out that in our two-qubit scheme, the success probability 25\% is for any arbitrary two-qubit states. Although in some previous CRSP schemes the success probabilities are higher than this, there were some restrictions on the state to be prepared, which make it not a really arbitrary state \cite{MCRSP_II_2GC3,CRSP_II_III_B}. In Ref. \cite{CRSP_I_A4} the success probability is 100\% for the CRSP scheme of an arbitrary single-qubit state. However, the receiver has to perform phase-shift operations, which requires the knowledge of the state to be prepared. Moreover, the controller has to perform the Bell state measurement, which makes the protocol more difficult to implement.

It is interesting to compare our schemes with previous CRSP protocols. On  one hand, the success probabilities of our two schemes are equal to that of schemes in which maximally entangled channels were employed \cite{MCRSP_I_NE_EPR,CRSP_II_E+G(C)}. On the other hand, all the previous schemes in which partially entangled channels are used require the assistance of auxiliary qubits and two-qubit unitary transformations \cite{CRSP_IIE_2WC3,MCRSP_II_2GC3,CRSP_II_E+G(C),CRSP_II_GC+WC1,CRSP_II_GC+WC2}, while our schemes need neither of them. Moreover, the success probabilities of these schemes depend on the parameters of the non-maximally entangled quantum channels while our schemes have certain success probabilities which are independent of the degree of freedom of the quantum channel.

Unlike our schemes which succeed probabilistically for arbitrary states, there is another class of schemes called controlled joint remote state preparation (CJRSP) schemes in which the arbitrary quantum state can be faithfully prepared with 100\% success probability. However, more quantum resources are required in those schemes. For example, an eight-qubit entangled quantum channel was utilized for preparation of arbitrary two-qubit states in Ref. \cite{CJRSP_II_CC8}. Each of the two senders performs two-qubit measurement according to the amplitude information and phase information of the state, respectively. And four classical bits should be transmitted to the receiver from the senders in total, which are the same as the classical information required in the teleportation scheme of two-qubit states \cite{ctele}. It is reasonable that more quantum resources bring the higher success probability and fidelity. In this paper, we just focused on the situation with only one sender.

Usually, the RSP schemes which use non-maximally entangled channels ask for auxiliary qubits to help eliminate the coefficients of the channel at the beginning or the end of the procedure \cite{CRSP_IIE_2WC3,MCRSP_II_2GC3,CRSP_II_E+G(C),CRSP_II_GC+WC1,CRSP_II_GC+WC2,CRSP_qudit2,CJRSP_II_CC8,CJRSP_I}. Then the total success probabilities are limited by the degree of entanglement of the quantum channels. However, the success probabilities of our schemes are independent of the quantum channel. This is due to  the interesting character of the maximal slice state, i.e., no matter what results the controller obtains, the state shared between the sender and the receiver collapsed onto the maximally entangled pair, which can be used for the remote state preparation with maximum success probability. The maximal slice state may have applications in other branches of quantum communication due to this feature.

In conclusion, although partially entangled channels are employed, our schemes require neither auxiliary qubits nor two-qubit unitary transformation, which makes our schemes more practical and resource-saving. The most important advantage of our schemes is that higher and independent success probabilities can be obtained by using non-maximally entangled channels, which will have good applications in quantum communication.

\begin{acknowledgements}
This work is supported by the National Natural Science
Foundation of China under Grant No. 11004258 and the Fundamental Research Funds for the Central Universities under
Grant No.CQDXWL-2012-014. It is also supported by the Natural Science Foundation Project of CQ CSTC 2011jjA90017.
\end{acknowledgements}

\end{document}